\documentclass[12pt]{article}
\usepackage[dvips]{graphicx}
\usepackage{amssymb}
\usepackage{amsmath}
\usepackage{epsfig}
\usepackage{cite}
\usepackage{hyperref}
\usepackage{bbold}
\usepackage{multirow}

\numberwithin{equation}{section}
\numberwithin{table}{section}

\def\beq{\begin{equation}}
\def\eeq{\end{equation}}
\def\be{\begin{equation}}
\def\ee{\end{equation}}
\def\bea{\begin{eqnarray}}
\def\eea{\end{eqnarray}}

\def\Im{{\rm Im\,}}

\DeclareRobustCommand{\SkipTocEntry}[4]{}
\RequirePackage{color}

\newcommand{\cO}{\mathcal{O}}

\newcommand{\dd}{\mathrm{d}}

\begin{document}

\begin{titlepage}
\begin{center}
\rightline{\small }

\begin{flushright} 
IPhT-T18/110 \\
CPHT-RR080.082018
\end{flushright}

\vskip 2cm

{\Large \bf Uplifting Runaways }
\vskip 1.2cm

{ Iosif Bena$^{*}$, Emilian Dudas$^{\dag}$, Mariana Gra\~na$^{*}$ and  
Severin L\"ust$^{*\, \dag}$ }
\vskip 0.1cm
{\small\it  $^{*}$ Institut de Physique Th\'eorique, 
Universit\'e Paris Saclay, CEA, CNRS\\
Orme des Merisiers \\
91191 Gif-sur-Yvette Cedex, France} \\
\vskip 0.1cm
{\small\it  $^\dag$ Centre de Physique Th\'eorique, Ecole Polytechnique, CNRS \\ 
91128 Palaiseau Cedex, France} \\
\vskip 0.8cm

{\tt }

\end{center}

\vskip 1cm

\begin{center} {\bf Abstract }\\

\end{center}

\vspace{0.2cm} 

We find a mechanism by which antibranes placed in a warped deformed conifold throat can destroy the stabilization of the size of the sphere at the tip, collapsing it to zero size. This {\em conifold destabilization mechanism} can be avoided by turning on a large amount of flux on the sphere, but tadpole cancelation makes this incompatible with a hierarchy of scales in a Type IIB flux compactification. This indicates that antibrane uplift cannot be used to construct stable de Sitter vacua with a small cosmological constant in perturbative String Theory. The values of $V$ and $V'$ for these KKLT-like scenarios can be parametrically small, but we find that $V'/V$ is still consistent with the de Sitter swampland conjecture. Our results also suggest that there should exist a Klebanov-Strassler black hole, holographically dual to a deconfined phase with spontaneously broken chiral symmetry.

\noindent

\vfill

\today

\end{titlepage}



\section{Introduction}

Obtaining de Sitter vacua in string compactifications has been a goal of string phenomenology ever since the discovery that the expansion of our universe is driven by vacuum energy. Since string theory produces naturally compactifications with zero or negative cosmological constant \cite{Maldacena-nogo}, there are two possible routes. The first is to add ``un-natural'' ingredients (such as smeared orientifold planes, non-geometric fluxes or T-branes) in the mix and to try to obtain de Sitter vacua as solutions of the full equations of motion. The second is to obtain these vacua indirectly, by first constructing supersymmetric anti-de Sitter vacua and then uplifting their cosmological constant to a positive one \cite{KKLT}. The second route offers much more technical control than the first, and can also produce de Sitter vacua whose cosmological constant is parametrically smaller than the string scale, which so far has proven impossible by direct construction.

However, uplifting does not come without problems. The most common method consists of adding anti-D3 branes in a flux compactification with stabilized moduli \cite{KKLT}. The uplift caused by the anti-D3 branes needs to be quite small, in order to ensure that the delicate mechanism responsible for the stabilization of the K\" ahler moduli is not disturbed. Hence the anti-D3 branes need to be placed in a region of high warping inside the compactification manifold. Even if the full solution for a warped compactification with a long warped throat has never been constructed, one usually assumes that the throat can be glued to a flux compactification and that the supersymmetry-breaking effects of the antibrane are localized near the bottom of this throat, and hence are very small from the perspective of the compactification scale. The prototypical example of a long warped throat is the Klebanov-Strassler warped deformed conifold \cite{KS}, and antibranes placed in this solution appear in the probe approximation to give rise to metastable vacua \cite{KPV}, whose supersymmetry-breaking effects are localized deep in the infrared. 

Over the past almost ten years, an extensive body of work has been devoted to ascertaining whether this probe-approximation picture is a correct description of the physics, or rather whether there is nontrivial physics that is missed by the probe approximation \cite{us,cloaking,tachyons,brotherhood-graveyard,thomas}. The overwhelming part of that work supports the second conclusion: antibranes source a singular solution, and their singularity cannot be cloaked by a black horizon. Furthermore, this singularity indicates the presence of a brane-brane repelling tachyon, which makes unstable any flux compactification constructed with antibranes. In the regime of parameters where the number of antibranes is small it was argued that the presence of a singularity can be understood in the framework of ``brane effective actions'' and that antibranes would eventually turn to be regular \cite{Joe-Andrea}; however, an explicit calculation of the brane effective action in a regime where exact calculations can be done shows that there are many unexpected cancelations \cite{Bena-Turton-Blaback}, and that multiple antibranes in Klebanov-Strassler throats are still unfeasible for obtaining de Sitter vacua by uplifting. 

Despite the multiple problems of antibranes revealed by the calculations of \cite{us,cloaking,tachyons,brotherhood-graveyard,thomas}, one can still hope that a single antibrane might still be usable for uplift. After all, a single antibrane does not suffer from a brane-brane-repelling tachyon, its worldvolume theory is free so the multi-loop calculations of  \cite{Bena-Turton-Blaback} do not apply to it, and its uplifting power is enough for obtaining de Sitter solutions with a small cosmological constant. A related construction, in which a single antibrane is placed on top of an orientifold three-plane, has also been argued to give an uplift mechanism \cite{kallosh}, and at the same time to be describable using a nilpotent supergravity action \cite{nilpotent,nilpotent-brotherhood-graveyard}. 

The main result of this paper is the discovery of a (previously ignored) {\em conifold destabilization mechanism}, which invalidates uplifting with both a single and with many antibranes. This mechanism does not give rise to an instability, but rather to a (much more pathological) runaway behavior. 

To understand this conifold destabilization mechanism one should remember that the Klebanov-Strassler solution is obtained by adding fluxes and warping to the deformed conifold, but one can also do that to the undeformed conifold (obtaining the singular Klebanov-Tseytlin geometry \cite{KT}). The deformation parameter of the conifold is a modulus, but in the infinite warped Klebanov-Strassler solution the warp factor ensures that the size of the three-sphere stays constant, and this modulus can be reabsorbed into an overall rescaling of the Minkowski coordinates. In an infinite Klebanov-Strassler solution this modulus is free to run, and one indeed expects it to run once supersymmetry-breaking ingredients are added. If one fixes a UV holographic cutoff, this running corresponds to changing the distance between the tip of the KS solution and the cutoff surface. Moreover, when the throat is glued into a flux compactification, this distance parameterizes the length of the so-called B-cycle, which is fixed by the fluxes. Indeed, a compactification with a KS throat is expected to have both the A cycle of the deformed conifold (threaded by M units of RR three-form flux) and the B cycle, threaded by NS-NS three-form flux, as well as other three-cycles.

It is also possible to see that as the size of the throat becomes longer and longer, the stabilization potential becomes shallower and shallower. This is not unexpected: the energy cost of deforming the B cycle at the bottom of a long throat, which can be quite large from the perspective of an observer living at the bottom of this throat, is warped away to a very small value from the perspective of an observer living up the throat. Since the superpotential is blind to warping effects, the effect of the warping on the stabilization potential can only come via corrections to the K\"ahler potential, and these corrections have been computed in \cite{Douglas:2007tu, Douglas:2008jx}. There is a crucial difference between the na\" ive stabilization potential, and the one computed taking the warping into account (both are displayed in Figure \ref{fig:VKS}). The correct potential has two minima, one in which the conifold deformation is finite, and one in which it is zero. The latter minimum corresponds to a Klebanov-Tseytlin (KT) throat. The na\" ive potential misses completely the minimum corresponding to the KT throat. 

Antibranes, or any other supersymmetry-breaking object, give rise to another term in the stabilization potential, equal to the energy of these objects. This energy depends on the position of these objects inside the compactification manifold, and is minimized when the antibranes are at the tip of the KS throat. The minimal energy is an increasing function of the conifold deformation parameter (depicted in Figure \ref{fig:VD3}). The (meta)stability, instability or runaway behavior of antibranes in long warped throats of flux compactification depends therefore on the interplay between this energy and the stabilization potential. Clearly, when the number of antibranes is large enough, they will  over-run the bump in Figure~\ref{fig:VKS}, and drive the throat towards the KT solution.

This conifold destabilization mechanism has important implications for two research area  where the KS solution plays an important role: String Phenomenology/Cosmology and Holographic QCD:

For flux compactifications, our analysis indicates that the deformation parameter of the conifold is very light and takes a special role among the complex structure moduli.
Therefore, it cannot be simply integrated out before considering the effect of antibranes.
A precise evaluation of all the factors that enter in the potential reveals that the stability of long warped throats depends on the interplay between the number of antibranes and the number of units of flux on the three-cycle of the deformed conifold, $M$.  At first glance this may look as encouraging news for the de Sitter landscape, but it is not. In order for the uplifting to work, the warping has to be very large, and this warping is controlled by the amount of NS-NS three-form flux on the B cycle, $K$. However, the RR and NSNS three-form fluxes give rise to a positive contribution to the D3 brane tadpole, equal to $M \times K$. This contribution needs to be canceled by the negative D3-charge of orientifold three-planes, and the largest known such contribution in type IIB compactification is $-32$. Hence, the only throats where antibranes are not giving rise to runaway behavior have a warping that is at most 1/5, which is much larger than the exponentially suppressed uplift term (of order $10^{-9}$ in \cite{KKLT}), needed to obtain a small cosmological constant and also to preserve the stabilization of the K\"ahler moduli. 

In parallel to our work there has been an extended effort, that is sometimes packaged under the ``weak gravity conjecture/swampland'' banner \cite{swampland-brotherhood-graveyard, Vafa}, to understand the kind of the effective theories that can be obtained as low-energy limits of fully consistent string theory models. The culmination of this bottom-up programme has been the recent conjecture by Obied, Ooguri, Spodyneiko and Vafa \cite{Vafa} that the de Sitter spaces obtained by controllable string-theory constructions are not only unstable (as the analysis of antibranes \cite{tachyons} would indicate), but have also runaways. In particular, it was argued that the ratio between the slope of the effective potential and its value is always bounded below by
$
{V' \over V} \ge a, 
$
where $a$ is a constant of order one in Planck units. The conifold destabilization mechanism we find is consistent with this runaway expectation. Moreover, our construction allows for the first top-down calculation of ${V' \over V}$ in a regime of parameters where $V$ is parametrically smaller than the string scale, and our results are consistent with the expectations of  \cite{Vafa}.

\subsection*{Holographic QCD}

The fact that antibranes or other supersymmetry-breaking objects can lead to a runaway of the KS solution  has also implications for strongly-interacting QCD-like physics. Indeed, the KS solution is the only known supergravity solution that is holographically dual to a supersymmetric confining gauge theory. As such, this solution is much better suited  for describing strongly-coupled quark-gluon-plasma physics than $AdS_5\times S^5$ (which is dual to a non-confining theory). 

A key question that one can hope to address using the KS solution is what is the nature of the deconfinement and the chiral symmetry breaking phase transitions, and whether they occur at the same temperature. The chiral symmetry is broken in the zero-temperature confining vacuum of this theory \cite{KS}, which is dual to the supersymmetric warped deformed conifold KS solution. The bulk solution dual to the high-temperature phase is a black hole in the warped undeformed conifold (Klebanov-Tseytlin) solution, and in this phase the chiral symmetry is restored  \cite{Aharony-Buchel-Kerner} (see \cite{KT-bh} for earlier work).

If the confinement and chiral symmetry breaking phase transitions occur at the same temperature, one only expects these two solutions to exist, but if the temperature of the chiral symmetry breaking phase transition is below that of the deconfinement phase transition, one expects to find a new solution, corresponding to a Klebanov-Strassler black hole. This solution would be dual to a phase with spontaneously broken chiral symmetry, but with deconfined quarks. An extensive numerical effort has gone towards the construction of this solution \cite{Buchel}, with no success. The only way to obtain a black hole solution dual to a phase with broken chiral symmetry has been to break this symmetry explicitly, by turning on a certain non-normalizable mode in the bulk (which forces the three-sphere at the bottom of the deformed conifold to stay large). Of course, the failure of a numerical construction can indicate either that the black hole does not exist, or that the numerics did not explore the full space of solutions. Our results support the second possibility, and suggest that a black hole in KS should exist. This solution would be dual to a deconfined phase with broken chiral symmetry, and if this phase is dominant this would indicate that the deconfinement and the $\chi$SB phase transitions happen at different temperatures. We will discuss this in more detail in Section \ref{Conclusions}.


\section{$\overline{D3}$-branes in a KS Throat}

\subsection{Calabi-Yau Manifolds with Fluxes and Warped Throats}
\label{step1}

The KKLT construction of de Sitter vacua \cite{KKLT} is based on a warped compactification on a Calabi-Yau manifold, with a constant dilaton, five and three-form fluxes \cite{GP}. 
The ten-dimensional metric and five-form flux are
\begin{equation}\begin{aligned}\label{eq:warpedbackground}
\dd s^2 &= H^{-1/2} \dd s^2_4 + H^{1/2} \dd s^2_{6} \,, \\
F_5 &= \left(1+\ast\right) \mathrm{vol}_4 \wedge \dd H^{-1} \equiv \ast {\cal F}_5+ {\cal F}_5 \,,
\end{aligned}\end{equation}
where $H$ is the warp factor and the $\dd s^2_{6}$ is the unwarped metric of the internal manifold. As argued in \cite{GKP}, one can think of this manifold as composed of a throat-type region of high warping, glued to a compact Calabi-Yau space. 
 
 In the region of high warping the local six-dimensional geometry is that of the deformed conifold, defined by its embedding into \(\mathbb{C}^4\),
\begin{equation}
\sum_{a = 1}^4 \omega_a^4 = S \,.
\end{equation}
The deformation parameter $S$ is the complex structure modulus whose absolute value corresponds to the size of the 3-sphere at the tip of the cone.
The other complex structure moduli $Z^I$ come from the ``UV" geometry. We thus have $h^{2,1}+1$ A-cycles:
\beq\label{eq:S}
\int_{A} \Omega_3= S \ , \qquad  \int_{A_I} \Omega_3= Z^I
\eeq
where $I=0,...,h^{2,1}-1$. We assume that the prepotential splits according to
\begin{equation}
F(S, Z^I) = F_{cf}(S) + F_{UV}(Z^I) \,,
\end{equation}
where $F_{cf}$ is the prepotential of the deformed conifold and  the ``UV prepotential,'' \(F_{UV}\), does not explicitly depend on \(S\).
We thus have
\beq\label{eq:GS}
\int_{B} \Omega_3 = F_S= \frac{S}{2 \pi i}\left( \log{\frac{\Lambda_0^3}{S}} +1 \right)+ F^{0}_S  \, , \qquad \int_{B_I} \Omega_3 = F_I
\eeq
where $F_S$  and $F_I\) are the derivatives of \(F\) with respect to $S$ and $Z_I$ respectively, and $F_S^{0}$ depends on the details of the compactification manifold, but is independent of $S$. The cutoff \(\Lambda_0\) corresponds to the transition between the highly warped region, modeled as a KS throat, and (relatively unwarped) rest of the compact Calabi-Yau manifold.
 
The 3-form fluxes on the 3-cycles are\footnote{The setup only requires one type of flux on each cycle.}
\begin{equation}\begin{aligned}\label{eq:fluxes2}
\frac{1}{(2\pi)^2 \alpha'} F_3 &= M \alpha + M^0 \alpha_0 - M_i \beta^i \,, \\
\frac{1}{(2\pi)^2 \alpha'} H_3 &= - K \beta  - K_0 \beta^0 + K^i \alpha_i  \,.
\end{aligned}\end{equation}
where $\alpha_i, \beta^i$ are Poincare duals to the cycles $B_i, A^i$ and we have singled out the RR flux on the $S^3$ cycle at the tip of the throat, $M$, and its NSNS partner $K$. These are the fluxes responsible for the deformation of the conifold by the parameter $S$, as we shall review.

The throat region is that of the Klebanov-Strassler solution  (KS) \cite{KS}, with the six-dimensional metric of the defomed conifold\footnote{Note that taking ${\cal T}$ and \(g^i\) to be dimensionless requires the deformation parameter \(S\) to be of dimension \((\mathit{length})^3\).}
\begin{equation}\begin{aligned}\label{eq:defconifoldmetric}
\dd s^2_6 = \frac{\left|S\right|^{2/3}}{2} \mathcal{K}({\cal T}) \biggl[\frac{1}{3 \mathcal{K}^3({\cal T})} \left(\dd {\cal T}^2 + (g^5)^2\right) + &\sinh^2({\cal T}/2)\left((g^1)^2 + (g^2)^2\right) \\ &+ \cosh^2({\cal T}/2)\left((g^1)^2 + (g^2)^2\right)\biggr] \,,
\end{aligned}\end{equation}
where \(g^i\) is an orthogonal basis of one-forms on the base of the cone and
\begin{equation}
\mathcal{K}({\cal T}) = \frac{\left(\sinh(2{\cal T}) - 2{\cal T}\right)^{1/3}}{2^{1/3}\sinh{\cal T}} \,.
\end{equation}
The warp factor of the KS solution is
\begin{equation}\label{eq:warpfactor}
H({\cal T})= 2^{2/3} \frac{{g_s (\alpha' M)^2}}{\left|S\right|^{4/3}} I({\cal T})
\end{equation}
where
\begin{equation}\label{eq:I}
I({\cal T}) = \int_{\cal T}^\infty \dd x \frac{x \coth x - 1}{\sinh^2 x} \left(\sinh(2x) -2x \right)^{1/3}.
\end{equation}
For large values of \({\cal T}\) one can introduce a more useful radial coordinate
\begin{equation}\label{eq:rtau}
r^2 = \frac{3}{2^{5/3}} \left|S\right|^{2/3} e^{2 {\cal T} /3} \,,
\end{equation}
 such that \eqref{eq:defconifoldmetric} approaches the conifold metric \(\dd r^2 + r^2 \dd s_{T^{11}}^2\).

Since the NSNS flux increases by $M$ units at every ``cascade" (borrowing  the dual gauge theory interpretation) the radial growth of the D3-charge dissolved in the fluxes is  $M^2 \, {\rm ln} (r)$. The UV cutoff $\Lambda_0$ where the solution is glued to the compact Calabi-Yau solution is such that the total NSNS flux over the $B$ cycle is $K$, according to \eqref{eq:fluxes2}:
\beq\label{eq:Kcutoff}
K= \frac{1}{(2\pi)^2 \alpha' } \int_{B} H_3=  \frac{1}{(2\pi)^2 \alpha' } \int_{{\cal T} \le {\cal T}_{0}} \int_{S^2} H_3  \ , \quad \ \Lambda_0^2=\frac{3}{2^{5/3}} \left|S\right|^{2/3} e^{2 {\cal T}_0 /3} \ . 
\eeq

On a compact manifold  the Bianchi identity for the five-form flux leads to the tadpole cancelation condition forcing the total D3-charge of the solution to be zero.
\beq \label{tadpole}
MK+M^0 K_0 - M_i K^i + Q^\mathrm{loc}_3= 0 \, .
\eeq
The charge of localized D3-brane and O3-plane sources is 
\beq
Q^\mathrm{loc}_3 = N_{D3}-\frac14 N_{O3} \, .
\eeq
On $T^6/Z_2$, where the $Z_2$ reverses the sign of all coordinates, there are $2^6=64$ O3-planes. On the other hand, if $M^I$, $K_I$ are those of the $T^6$ without the $Z_2$ action, there is an extra factor of 2, and the tadpole condition is
\beq \label{tadpoleT6}
MK+M^0 K_0 - M^i K_i =-Q^\mathrm{loc}_3 =32-2 N_{D3} + 2 N_{\overline{D3}}\ .
\eeq
Although nobody has demonstrated the existence of an upper bound for the number of orientifold planes, or rather the number of possible $Z_2$ actions with fixed points, on a given Calabi-Yau manifold\footnote{We thank Andre Lukas, David Morrison, Fabian Ruehle and Timo Weigand for discussions on this point.}, in all the examples used in the literature on moduli stabilization by fluxes, the number of O3-planes is always less or equal than 64. We will assume then that the tadpole cancelation condition is at best of the form \eqref{tadpoleT6}. Of course, here we are restricting to compactifications that have a string theory perturbative description.  
In F-theory compactifications there is a contribution to the tadpole from wrapped (p,q) 7-branes. This is nicely encoded geometrically by the Euler number of the elliptically fibred four-fold, which has no upper bound and can give a much higher contribution than \eqref{tadpoleT6}.

\subsection{The Flux-induced Potential}
\label{sec:vks}

The fluxes \eqref{eq:fluxes2} give rise to a potential that generically fixes the complex structure moduli $S, z^i$ and the dilaton \(\tau = C_0 + \ i e^{-\Phi}\).
This potential comes from the Gukov-Vafa-Witten superpotential \cite{GVW} 
\begin{equation} \label{WGVW} \begin{aligned}
\frac{1}{(2\pi)^2 \alpha'} W &=  \frac{1}{(2\pi)^2 \alpha'} \int (F_3-\tau H_3) \wedge \Omega \\
&=-M F_S - \tau KS - M^0 F_0 +M_i Z^i - \tau \left( K_0 Z^0 - K^i F_i\right) . \\
\end{aligned}\end{equation}

\begin{figure}[htb]
\centering
\includegraphics[width=10cm]{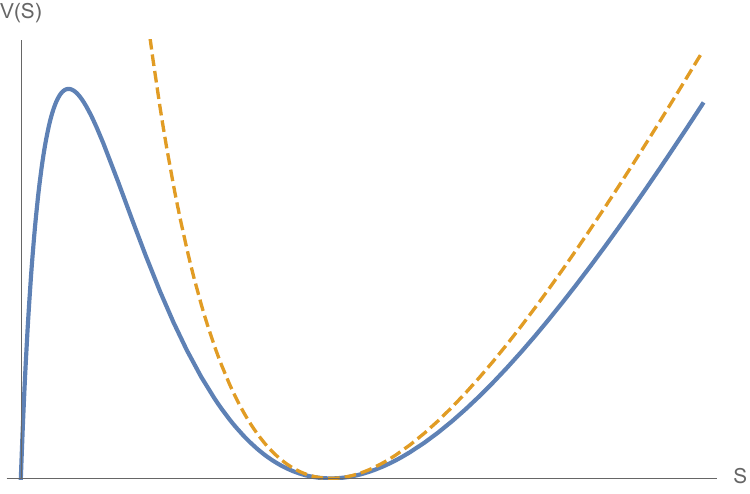}
\caption{The potential \(V_{KS}\) of \cite{Douglas:2007tu} for the complex structure modulus \(S\) of the Klebanov-Strassler throat given in \eqref{eq:VKS}.
The solid blue line corresponds to the full potential, while the dotted orange line does shows the na\" ive potential that does not take into account  the effects of warping (\(c' = 0\)).
Both potentials have the same supersymmetric minimum but differ drastically at small $S$.}
\label{fig:VKS}
\end{figure}

The potential for the complex structure modulus $S$ involves the fluxes $M$ and $K$, while it depends on the other fluxes only indirectly through the axion-dilaton $\tau$, whose vev is determined by all fluxes. Furthermore, unlike the other ``bulk" moduli, the potential for $S$ is highly affected by the warp factor.
Its functional form, derived in \cite{Douglas:2007tu, Douglas:2008jx} is\footnote{We follow the Einstein frame conventions of \cite{GKP} and use $2 \kappa_{10}^2 = (2 \pi)^7 \alpha'^4$.} 
\begin{equation}\label{eq:VKS}
V_{KS} = \frac{\pi^{3/2}}{\kappa_{10}} \frac{g_s}{(\Im\rho)^3}\left[c \log\frac{\Lambda_0^3}{\left|S\right|} + c'\frac{{g_s (\alpha' M)^2}}{\left|S\right|^{4/3}}\right]^{-1} \left|\frac{M}{2 \pi i} \log\frac{\Lambda_0^3}{S} + i \frac{K}{g_s} \right|^2  \,,
\end{equation}
where $g_s$ is the stabilized vev of the dilaton, $\Im \rho=({\rm Vol}_6)^{3/2}$ (see Appendix \ref{sec:4D} for more details),  
\(c\) denotes the constant value of the warp factor at the UV and will not be relevant here, whereas the constant \(c'\), multiplying the term coming solely from the warp factor, denotes an order one coefficient, whose approximate numerical value was determined in \cite{Douglas:2007tu} to be
\begin{equation}
c' \approx 1.18 \,.
\end{equation}
The potential \(V_{KS}\) is plotted in Figure~\ref{fig:VKS}.

The potential \eqref{eq:VKS} has a supersymmetric minimum, corresponding to \(\partial_S W = 0\), which, for $S\ll\Lambda_0^3$, is at
\begin{equation} \label{sKS}
s_\mathrm{KS} \simeq  \Lambda_0^3 \exp\left(-\frac{2\pi K}{g_s M}\right) \, .
\end{equation}
We define the quantity $h$, which measures the hierarchy between the UV and IR scales 
\beq \label{hierarchy}
h=3 \ln \frac{\Lambda_0}{\Lambda_{\rm{IR}}} =\frac{2 \pi K}{g_sM} \ .
\eeq

It is instructive to compute also the mass of $S$ at this minimum.
It can be obtained from
\begin{equation}
m^2_S \equiv \frac{1}{M_{pl}^2} G^{S \bar S} \partial_{\bar S} \partial_S V \Bigr|_{S = s_\mathrm{KS}} \,.
\end{equation} 
The metric \(G_{S \bar S}\), which is used to obtain \(V_{KS}\) in \eqref{eq:VKS} in the first place, is given in \cite{Douglas:2007tu}:
\begin{equation}\label{eq:Smetric}
G_{S \bar S} = \frac{1}{\pi \left\|\Omega\right\|^2 V_{w}} \left(c \log\frac{\Lambda_0^3}{\left|S\right|} + c'\frac{{g_s (\alpha' M)^2}}{\left|S\right|^{4/3}}\right) \,.
\end{equation}
Here \(\left\|\Omega\right\|^2 = \Omega_{ijk} \bar \Omega^{ijk}/3!\) and \(V_w = \int \sqrt{g_6}  H\) is the warped volume \cite{DeWolfe:2002nn}, with $H$ the warp factor of the ten-dimensional solution \eqref{eq:warpedbackground}. Thus
\begin{equation}\label{eq:OmegaVw}
\left\|\Omega\right\|^2 V_w = \int H \, \Omega \wedge \bar \Omega \,.
\end{equation}
Notice that \(\left\|\Omega\right\|^2 V_w\) drops out of the potential itself.
For small values of $|S|$ the second term in \eqref{eq:Smetric}, which is proportional to $c'$, dominates.
This term accounts for the effects of the warp factor of the Klebanov-Strassler solution.
Hence, in the regime where $S = s_{KS}$ is small we can effectively set $c = 0$ and obtain 
\begin{equation}
m^2_S = \frac{\pi^4\left\|\Omega\right\|^2}{c'^2 g_s M^2}  \frac{s_{KS}^{2/3}}{\alpha'^2} \,.
\end{equation}
On the other hand, if one ignores the contributions of the warp factor to the metric by setting $c'=0$ one finds 
\begin{equation}
m^2_S = \frac{\pi^2 \left\|\Omega\right\|^2 g_s^3 M^4}{4 c^2 K^2} \frac{\alpha'^2}{s^2_\mathrm{KS}} \,.
\end{equation}
A similar result has been obtained in \cite{Blumenhagen}, where it was concluded that the mass of $S$ is thus exponentially large.
However, as we have seen before, including the effects of the warping gives a qualitatively different result.
This is consistent with our following analysis where we will show that $S$ is so light that the addition of an anti-D3-brane has a non-negligible effect on its stabilization.
 
Let us finally study the behavior of \eqref{eq:VKS} in the decompactification limit towards an infinitly long Klebanov-Strassler throat.
To perform this limit in a controlled way we should keep in mind that \eqref{eq:Kcutoff} implies that the flux $K$ along the $B$-cycle grows logarithmically with $\Lambda_0$.
Therefore, one should not send only $\Lambda_0$ to infinity but take instead the limit
\begin{equation}\label{eq:infinitlimit}
\Lambda_0, K \rightarrow \infty \,\qquad\text{with}\qquad \Lambda_0^3 \exp\left(-\frac{2\pi K}{g_s M}\right) = \mathrm{fixed} \,,
\end{equation}
such that the minimum $s_{KS}$ in \eqref{sKS} stays always at the same position.
So if one chooses initial values for $\Lambda_0$, $K$ and $M$ such that $s_{KS}$ is small, it will remain small even if the throat becomes extremely long and we are still in the regime of validity of the potential $V_{KS}$ (\eqref{eq:VKS}).
To understand the behavior of $V_{KS}$ in the limit \eqref{eq:infinitlimit}, we first notice that its numerator $\sim \left|\partial_S W\right|^2$ depends on $\Lambda_0$ and $K$ exlusively via their combination in $s_{KS}$.
Hence it does not change under \eqref{eq:infinitlimit}.
On the other hand, the denominator of \eqref{eq:VKS} does not explicitly depend on $K$, but blows up for $\Lambda_0 \rightarrow \infty$ and fixed $S$:
\begin{equation}
c \log\frac{\Lambda_0^3}{\left|S\right|} + c'\frac{{g_s (\alpha' M)^2}}{\left|S\right|^{4/3}}  = c \log \Lambda^3_0 + \mathcal{O}(K^0 \Lambda_0^0) \rightarrow \infty \,,
\end{equation}
under \eqref{eq:infinitlimit}.
Therefore, \(V_{KS}\) converges pointwise to a flat potential:
\begin{equation}
V_{KS}(S) \rightarrow 0 \qquad\text{for}\qquad \left|S\right| > 0 \,.
\end{equation}
This confirmes the intuition that \(S\) is an exact modulus of the infinite Klebanov-Strassler solution that can be varied without any cost in energy.

\subsection{The potential of an anti-$D3$ brane}
\label{sec:step3}

An anti-D3 at the tip of the throat uplifts the KS potential \eqref{eq:VKS}. The contribution to the potential is determined from
\begin{equation}\begin{aligned}\label{eq:D3action}
S_{D3} = S_{DBI} + S_{CS}
= - T_3 \int \dd^4 x \sqrt{-g_4} \bigl[1 + \cO(\alpha'^2)\bigr] \pm T_3 \int C_4 \,,
\end{aligned}\end{equation}
where the sign in front of the second term is determined by the charge of the brane, and 
 \(T_3\) is given by
\begin{equation}
T_3 = \frac{1}{(2 \pi)^3 \alpha'^2} \,.
\end{equation}
It is not hard to see that for the D3-brane in a background given by \eqref{eq:warpedbackground}, the DBI and the CS pieces of the action cancel each other. Hence, for the \(\overline{\mathrm{D3}}\)-brane they add up and one finds
\begin{equation}
V_{\overline{D3}} = - 2 T_3 C_4 = \frac{2}{(2 \pi)^3 \alpha'^2} H^{-1} \,.
\end{equation}
Using the warp factor given in \eqref{eq:warpfactor} we finally obtain
\begin{equation} \label{VantiD3}
V_{\overline{D3}} = \frac{ \pi^{1/2}}{\kappa_{10}}  \frac{1}{(\Im\rho)^3}  \frac{2^{1/3}}{I(\tau)} \frac{\left|S\right|^{4/3}}{g_s (\alpha' M)^2} \,.
\end{equation}
Because \(I(\tau)\), defined in \eqref{eq:I}, is a monotonically decreasing function, this expression has a minimum at \(\tau = 0\).
Consequently, a \(\overline{\mathrm{D3}}\)-brane has minimal energy if it is placed at the tip of the throat. For later convenience we introduce
\begin{equation}
c'' = \frac{2^{1/3}}{I(0)} \approx 1.75 \,.
\end{equation}
For $N$ anti-D3 branes the potential is multiplied by $N$, and this is taken care by simply replacing $c''\to c''N$.

\begin{figure}[htb]
\centering
\includegraphics[width=10cm]{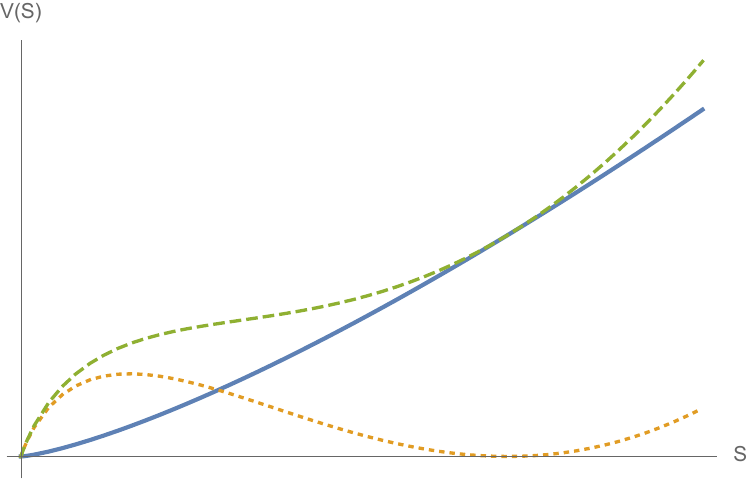}
\caption{The contribution \(V_{\overline{D3}}\) (solid blue line) of an \(\overline{D3}\)-brane placed in the Klebanov-Strassler throat to the potential for \(S\).
The two other lines represent the original potential \(V_{KS}\) (dotted orange line) for the specific value \(\sqrt{g_s} M = 6\) as well as the superposition \(V_{KS} + V_{\overline{D3}}\) (dashed green line).} 
\label{fig:VD3}
\end{figure}

In Figure~\ref{fig:VD3} we plot the \(\overline{D3}\)-potential \(V_{\overline{D3}}\) together with the superposition of \(V_{KS}\) with \(V_{\overline{D3}}\).

\section{The Total Potential}

\begin{figure}[htb]
\centering
\includegraphics[width=10cm]{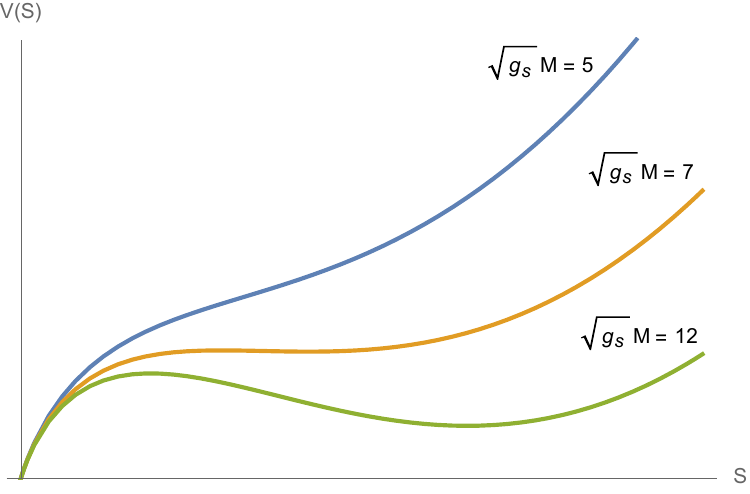}
\label{fig:VKSD3}
\caption{The combined potential \(V_{KS} + V_{\overline{D3}}\) for one anti-D3 brane and \(\sqrt{g_s}M = 5\), 7 and 12.
All three graphs are drawn for the same ratio \(K/M = 5\).
A local minimum only exists if \(M\) is larger than the threshold value \(M_\text{min} \approx 6.8\).
}
\end{figure}

We consider the combined potential \(V_{KS} + V_{\overline{D3}}\) as illustrated in Figure~\ref{fig:VD3}.
The analysis simplifies considerably if one neglects the first term in the denominator of \(V_{KS}\) in \eqref{eq:VKS}, which 
can safely be done as long as
\begin{equation}\label{eq:mincondition}
\frac{{g_s (\alpha' M)^2}}{\left|S\right|^{4/3}} \gg \log\frac{\Lambda_0^3}{\left|S\right|} \,.
\end{equation}
In this regime of parameters \(V_{\overline{\mathrm{D3}}}\) gives only a constant contribution to the numerator of \(V_{KS}\) and the critical points of the combined potential \(V_{KS} + V_{\overline{D3}}\) can be determined analytically. 
They are given by
\begin{equation}\label{eq:smin}
s_\mathrm{crit} = \Lambda_0^3 \exp\left(-\frac{2\pi K}{g_s M} -\frac{3}{4} \pm \sqrt{\frac{9}{16} - \frac{4 \pi}{g_s M^2} c' c''}\right) \,,
\end{equation}
whereas the positive sign correspondents to a local minimum and the negative sign to a local maximum.
Thus, the total potential for $N$ anti-D3 branes has extrema only for\footnote{
The factor of $\sqrt{g_s}$ was missing in the first version of this paper and has been corrected in \cite{Blumenhagen:2019qcg}.
We thank Ralph Blumenhagen for correspondence regarding this point.
We furthermore corrected the numerical value of $M_\mathrm{min}$ with respect to the first version.} 
\begin{equation}\label{eq:Mconstraint}
\sqrt{g_s} M > M_\mathrm{min} \,\qquad\text{with}\qquad M_\mathrm{min} = \frac{8}{3} \sqrt{ \pi c' c''} \approx 6.8 \sqrt{N}\,.
\end{equation}
Otherwise the potential becomes monotonically increasing and the only minimum lies at \(s = 0\). 
This is illustrated in Figure~\ref{fig:VKSD3}, where we plot the combined potential for different values of \(\sqrt{g_s} M\)  for a single anti-D3 brane, which we restrict to from now on since it gives the least strong constraint on $\sqrt{g_s} M$.
As we will show, this minimum value for $\sqrt{g_s} M$ is in strong tension with the tadpole cancelation condition and the requirement of a large hierarchy. 

\subsection{de Sitter minima and hierarchy}
\label{dS-hierarchy}

Requiring the potential to have a critical point forces the lower bound $\sqrt{g_s} M \gtrsim 6.8$.
On the other hand, there is another bound on $M K$ from above by the tadpole cancelation condition \eqref{tadpole}
\begin{equation}\label{eq:Kbound}
M K \le \left|Q_3^\mathrm{loc}\right| \,.
\end{equation}
Of course, this bound can only be saturated if there is one complex structure modulus since the flux required to stabilize additional moduli would contribute to the tadpole cancellation condition as well (moreover, the bound might not be integer).
Inserting both bounds in the quantity $h$ defined in \eqref{hierarchy}, which determines the hierarchy of scales, gives
\begin{equation}\label{boundhierarchy}
h = 2 \pi \frac{M K}{g_s M^2} <  2\pi \frac{\left|Q_3^\mathrm{loc}\right|}{M^2_\mathrm{min}} \approx 4.6
  \,.
\end{equation}
where the estimation above uses
the value of $|Q_3^{loc}| = 32+2$ for an orientifold of a torus and one anti-D3 brane. 
We therefore conclude that in a compactification where antibranes do not cause throat runaways 
\beq
\frac{\Lambda_{IR}}{\Lambda_{UV}} > 0.2 \ ,
\eeq
which cannot give rise to a de Sitter vacuum with a hierarchy of scales

\subsection{Conifold Runaways and de Sitter Swamplands}

It was recently conjectured by Obied, Ooguri, Spodyneiko and  Vafa \cite{Vafa} that in all controlled compactifications 
\begin{equation}\label{eq:VprimeoverV}
\left|\nabla V\right| \geq a V \,,
\end{equation}
for some constant $a$ of order one in Planck units. This would exclude the existence of de Sitter vacua in perturbative string theory.

In the previous section we have found that the existence of a de Sitter vacuum in our setup depends crucially on the amount of flux \(M\) along the A-cycle of the Klebanov-Strasler throat.
Hence, we want to evaluate \(\left|\nabla V\right|/V\) for different values of \(M\).

Assuming that all other complex structure moduli are stabilized and neglecting the contribution of any K\"ahler moduli the expression we will compute is given by 
\begin{equation}
\frac{\left|\nabla_S V\right|}{V} = \frac{\sqrt{G^{S\bar S} \partial_S V \partial_{\bar S} V}}{V} \,.
\end{equation}
with the metric $G_{S\bar S}$ given in \eqref{eq:Smetric}.

As before we assume that \(|S|\) is small enough such that the second term in \eqref{eq:Smetric} dominates.
In this limit we obtain\footnote{The warped volume appearing in the inverse metric drops out in the conversion of string units into Planck units.} 
\begin{equation}\label{eq:VprimeoverVresult}
\frac{\left|\nabla_S V\right|}{V} = \frac{\pi^{1/2} \left\|\Omega\right\| V_w^{1/2}}{\sqrt{c'} \alpha' \sqrt{g_s} M} \frac{\bigl|f(|S|)\bigr|}{|S|^{1/3}} \,,
\end{equation}
where we have introduced
\begin{equation}
f(s) = \frac23 + \left(2 \pi K + g_s M \log\frac{s}{\Lambda^3_0}\right)\left[\frac{4 \pi}{g_s M} (\pi K^2 + c' c'' g_s) +  \log\frac{s}{\Lambda^3_0} \left(4 \pi K + g_s M \log\frac{s}{\Lambda^3_0}\right)\right]^{-1} \,.
\end{equation}
This function has a minimum at
\begin{equation}
s = \exp\left[- \frac{2 \pi K}{g_s M} \left(1 + \frac{\sqrt{c' c'' g_s} }{\sqrt{\pi}K}\right)\right] \,,
\end{equation}
where it takes the value
\begin{equation}
f_\text{min} = \frac23 \left(1 - \frac{\sqrt{g_s} M}{M_\text{min}}\right) \,,
\end{equation}
with \(M_\text{min}\) as in \eqref{eq:Mconstraint}.
Moreover, the asymptotic behavior of \(f(s)\) is given by
\begin{equation}
f(s \rightarrow 0) = f(s \rightarrow \infty) = \frac23 \,,
\end{equation}
so \(f_\text{min}\) is a global minimum and bounds \(f(s)\) from below.

For \(\sqrt{g_s} M > M_\text{min}\) we find that \(f_\text{min}\) is negative and thus \(f(s) = 0\) for some value of \(s\).
Thus, the complex-structure modulus $S$ is stabilized and the conjecture \eqref{eq:VprimeoverV} does not appear to hold along this direction. However, for this value of $M$ there is no hierarchy of scales, and hence the resulting cosmological constant is string scale. Furthermore, such a large uplift will probably over-run the stabilization of the K\"ahler moduli, giving rise to other runaway behaviors.

On the other hand, if \(\sqrt{g_s}M < M_\text{min}\) we find
\begin{equation}\label{eq:VprimeoverVestimate}
\frac{\left|\nabla_S V\right|}{V} \geq \frac{2 \pi^{1/2} \left\|\Omega\right\| V_w^{1/2}}{3 \sqrt{c'} \alpha' \sqrt{g_s} M} \frac{1}{|S|^{1/3}}  \left(1 - \frac{\sqrt{g_s}M}{M_\text{min}}\right) \,.
\end{equation}
To bring this into a less obscure form one has to canonically normalize the field $S$.
From \eqref{eq:Smetric} we know that its kinetic term is
\begin{equation}
\frac{1}{\kappa^2_4} G_{S\bar S} \left|\dd S\right|^2 = M^2_{pl} \frac{ c' }{\pi \left\|\Omega\right\| V_w^{1/2}} \frac{{g_s (\alpha' M)^2}}{\left|S\right|^{4/3}} \left|\dd S\right|^2 \,,
\end{equation}
where we have again neglected the first term of $G_{S \bar S}$ which is subleading as long as $\left|S\right|$ is small.
Consequently, we should introduce another field $\phi$ defined by
\begin{equation}
\phi = \frac{3 M_{pl} \sqrt{c'}}{\pi^{1/2} \left\|\Omega\right\| V_w^{1/2}} \alpha' \sqrt{g_s} M S^{1/3} = \frac{3 \sqrt{g_s} M \sqrt{c'}}{8 \pi^4 \alpha' \left\|\Omega\right\|} S^{1/3} \,,
\end{equation}
which has mass dimension one and a canonical kinetic term.
In the second step we used $M^2_{pl} = \kappa^{-2}_4 = V_W \kappa^{-2}_{10}$.
Moreover, rewriting \eqref{eq:VprimeoverVestimate} in terms of $\phi$ shows that
\begin{equation}
\frac{\left|\nabla_\phi V\right|}{V} \geq \frac{2 M_{pl}}{\left|\phi\right|} \left(1 - \frac{M}{M_\text{min}}\right) \,.
\end{equation}

We would like again to emphasize that this calculation is done assuming a long throat and a very large hierarchy before the addition of the antibrane. Hence, both $V$ and $V'$ can be arbitrarily small in Planck units (most of the other estimates of $V'/V$ have been done in the regime of parameters where $V$ is large). The fact that the conjecture \eqref{eq:VprimeoverV} is still satisfied is a nontrivial confirmation.

\section{Conclusions and Future Directions}
\label{Conclusions}

Our calculation indicates that adding antibranes to warped throats glued to Type IIB flux compactications gives rise to a runaway behavior. 
The origin of this runaway is the fact that the deformation parameter of the conifold is a very light field. The usual assumption that all complex structure moduli can be integrated out before adding an anti-D3 brane should thus be revised with care.

There are three ways to avoid the runaway behavior:
\smallskip
\\
$\bullet$ The first is to work in F theory, where the D3 brane tadpole is canceled by flux on 7-branes whose negative D3-charge is encoded in the Euler number of the Calabi-Yau 4-fold. These Euler numbers can be quite large, so these compactifications could in principle accommodate a throat with a large $M$ and a lot of warping. The price is that from a string theory perspective one loses all perturbative control. \smallskip \\
$\bullet$ The second is to try to construct stabilized de Sitter compactifications using huge uplift terms. These compactifications will generically have cycles of order the string length, and thus these constructions have no perturbative control either. \smallskip \\
$\bullet$ The third is to use objects in string theory that are lighter than a single D3 brane. The first object one may think of is ``half'' an anti-D3 brane stuck to an orientifold plane \cite{uranga}, whose contribution to the effective potential is half of that of a regular anti-D3 brane. Blindly adding such a contribution to the potential of the stabilized warped deformed conifold throat will not help\footnote{It will decrease the minimum value of $\sqrt{g_s} M$ by a factor of $\sqrt{2}$, and increase 
$\Lambda_{IR}/\Lambda_{UV}$ from $1/5$ to $1/25$.}. Adding an orientifold plane at the bottom of a warped throat \cite{madrid} will probably also modify some terms in the stabilization potential, and it would be interesting to accurately calculate the new upper bound on the hierarchy, but we do not expect it to be much different from the one we find.

\subsection*{Klebanov-Strassler black holes}

As we discussed in the Introduction, our results suggest that, despite the failure of numerics to find a KS black hole \cite{Buchel}, such a black hole could exist in a certain temperature range. To see this, one should remember the technique used in the numerical construction of \cite{Buchel}: the putative KS  black hole solution is parameterized by nine functions of one variable, satisfying coupled second-order differential equations. There are boundary conditions in the UV and at the black hole horizon, which ensure that the solution is physical and does not have non-normalizable modes. After fixing these conditions one is left with several arbitrary constants both in the UV and in the IR expansions, and one finds these constants by shooting the eight functions both from the UV and from the IR, and making sure that they match in the middle. The only solutions produced by this procedure were the KT black hole as well as black holes where the chiral symmetry was broken explicitly by turning on a non-normalizable mode. 

Since the numerics were done in the infinite KS solution, the potential $V_{KS}$ that stabilizes the modulus corresponding to the deformation of the conifold (given in equation \eqref{eq:VKS} and discussed in detail in Section \ref{sec:vks}) is zero. Hence one may expect that any non-trivial amount of energy, either coming from an antibrane or from a non-extremal black hole horizon, will drive the conifold deformation to zero and bring one to the KT solution. This can be seen explicitly by taking the infinite throat limit \eqref{eq:infinitlimit} of the full $V_{KS}+ V_{\overline D3}$ potential. Since $V_{KS}$ vanishes in this limit while $V_{\overline D3}$ survives, the deformation of the conifold will be driven to zero by the addition of any energy source. This appears to support the non-existence of KS black holes.

However, the physics is a bit more subtle. In particular, when taking the infinite throat limit, the energy of an antibrane evaluated at the scale of the holographic screen also becomes vanishingly small, and one may hope that the vanishing of this energy contribution is comparable to the vanishing of $V_{KS}$, such that the full potential still has a metastable minimum. This can be seen explicitly by working in the regime of parameters given by equation  \eqref{eq:mincondition}: in this regime the existence of a local minimum with antibranes is guaranteed as long as the number of antibranes one adds is less than $g_s (M/6.8)^2$, independent of the length of the throat !  If, instead of antibranes, one tries to add a nonextremal black hole horizon at a finite distance away from the tip of the deformed conifold, we also expect a metastable minimum, and hence a KS black hole.

Thus, the question of whether the KS black hole exists boils down to the suitability of the UV gluing of the KS solution to a compact CY to act as a holographic screen. Clearly, such a black hole exists when the holographic cutoff, $\Lambda_0$, is in the regime \eqref{eq:mincondition}, which contains a very large number of duality cascades in the dual gauge theory. Moreover, if the holographic correspondence is correct, we do not expect the (infrared) physics of confinement and chiral symmetry breaking (an hence the existence of a KS black hole) to depend on the location of the UV holographic screen. Thus, the only possibility consistent with holography is that gluing a KS throat to a compact CY is a correct model for a UV holographic screen only in the regime of parameters \eqref{eq:mincondition}. The independence of the IR physics on the location of the holographic screen therefore supports the existence of a KS black hole.

The mass above extremality brought about by the black hole gives rise to a term in the energy that has exactly the same dependence on $S$ as the antibrane energy (\ref{VantiD3}). Hence, if the value of $M$ is large we expect to find a KS black hole, whose mass is bounded above by a term proportional to $g_s M^2$. Of course, since this black hole is smeared over the $S^3$, whose radius scales like $M^{1/2}$, the horizon radius (and hence the inverse temperature) would scale like $\sqrt{M}$. Thus, when $M$ is very large, we expect to find KS black holes at temperatures higher than $1/\sqrt{M}$. 

It is clearly very important to try to construct this KS black hole, and thus confirm our prediction. The existence of this black hole would indicate that the confining gauge theory dual to the KS solution has a new intermediate-temperature phase, where chiral symmetry is broken but the quarks are not confined. Its numerical construction would allow us to ascertain the regime of parameters where this new phase dominates, and we hope the discovery of this new phase to lead to new lattice investigations and a deeper understanding of the deconfinement and chiral symmetry breaking phase transitions of strongly-coupled quark-gluon plasmas. \\

\noindent {\bf Note added: }  
After this paper was submitted to the arXiv, our prediction for the existence of a  KS  black hole was confirmed by the numerical construction of this black hole in  \cite{Buchel:2018bzp}. \\

\noindent {\bf Acknowledgements:} 
We would like to thank S. Massai for useful early discussions on this topic. We have also benefited from interesting discussions with E. Iancu, A. Lukas, J. Maldacena, B. Michel, A. Micu, D. Morrison, A. Puhm, F. Ruehle, C. Vafa and T. Weigand. We thank R. Blumenhagen for correspondence regarding a missing factor of $g_s$ in some of the equations in the first version of this paper. This work was supported in part by the ANR grant Black-dS-String ANR-16-CE31-0004-01 and the ERC Consolidator Grant 772408-Stringlandscape.
 

\newpage

\appendix
\noindent
{\bf\Huge Appendix}

\section{Four-dimensional supergravity description}\label{sec:4D}

It is instructive to translate the 10d description of the conifold destabilization mechanism to the language of four-dimensional supergravity. Before the uplift, the 4D description of the Calabi-Yau compactification with O3 planes and fluxes is given by $N=1$ supergravity with $1+h^{2,1}+h^{1,1}$ chiral multiplets corresponding respectively to the axion-dilaton, the complex structure moduli measuring sizes of 3-cycles and the K\"ahler moduli, which are a combination of the size of the 4-cycles and the RR 4-form potential.  

Three-form fluxes generate a superpotential for the complex structure moduli and the dilaton, which are stabilized in a supersymmetric Minkowski vacuum. The K\"ahler moduli are not fixed by the fluxes, and have moreover a runaway potential as \eqref{eq:VKS} shows. However, the presence of Euclidean D3-brane instantons (wrapping four-cycles of the Calabi-Yau manifold) or gaugino condensation on D7-branes stretched along four-cycles generate a superpotential for  these moduli at the non-perturbative level. The details of this mechanism are not relevant here, and some problems with it have been highlighted in \cite{sav-ander}, but we will show that even if the KKLT K\"ahler moduli stabilization mechanism still works, it will not help in avoiding the problem we found.

Assuming there is only one K\"ahler modulus, $\rho$, the potential has a supersymmetric AdS minimum at
\beq
V_{{\rm AdS}}= -\frac{a^2 A^2 e^{-2a \Im \rho_0}}{\Im \rho_0}  \ , \label{4d1}
\eeq
where $A$ and $a$ are parameters in the non-perturbative superpotential\footnote{The parameters $A$ and $a$ are not constants but depend actually on the complex structure and dilaton moduli, but since these are fixed by a perturbative mechanism, their masses are assumed to be large enough to decouple from the low energy effective theory for the K\"ahler modulus.}, and $\rho_0$ is a solution to 
\beq
W_0=-A e^{-a \Im\rho_0} \left(1+\frac23 a \Im \rho_0\right)  \ , \label{4d2}
\eeq
where $W_0$ is the flux-generated superpotential \eqref{WGVW} evaluated at the values of the dilaton and complex structure moduli that minimize the perturbative potential. In order for the uplift mechanism to give a positive and small cosmological constant, one needs to fine-tune $W_0$ to a very small value.

Finally, the uplift can also be described in a manifestly supersymmetric formalism using nonlinear supersymmetry with a nilpotent goldstino superfield \cite{nilpotent, nilpotent-brotherhood-graveyard,kallosh}.  Therefore, if there is a mass gap we should be able to describe the whole action in terms of a supergravity action. Introducing a nilpotent superfield $X$, it is indeed possible to write the K\"ahler potential at the perturbative level as
\begin{equation}\begin{aligned}
K &= - 3 \log \left(-i (\rho - \bar \rho)\right) -\log\left(- i (\tau - \bar \tau)\right) - \log\left(\frac{\left\|\Omega_0\right\|^2 V_w^2}{\kappa^{12}_4}\right) \\
 &\qquad + \frac{1}{\pi \left\|\Omega_0\right\|^2 V_w} \left[c |S|^2 \left(  \log  \frac{\Lambda_0^3}{|S|} + 1 \right)  + 9 c'  g_s (\alpha' M) ^2 |S|^{2/3}  +  |X|^2 \right] \,,\\
\end{aligned}\end{equation}
where the warped volume $V_w$ and $\left\|\Omega\right\|^2$ are defined above \eqref{eq:OmegaVw}.
Here, $\left\|\Omega_0\right\|^2$ denotes the value of $\left\|\Omega\right\|^2$ at fixed $S$,
such that $K$ depends on $S$ only explicitly via the term in the second line.  
This K\"ahler potential should be understood as an $S$-expansion of the general K\"ahler potential derived in \cite{DeWolfe:2002nn}, reproducing the metric $G_{S \bar S}$ of \cite{Douglas:2007tu, Douglas:2008jx}, given in \eqref{eq:Smetric}.
On the other hand, assuming the dilaton and all other complex structure moduli besides $S$ are fixed, the superpotential is%
\footnote{Note that this expression differs by a factor of $\kappa_4^{-8}$ from the superpotential used in the main part of this paper.
This change is necessary to obtain a superpotential of the conventional mass dimension in four-dimensional supersymmetry.} 
\begin{equation}
\frac{\kappa_4^8}{(2 \pi)^2 \alpha'} W = \frac{M}{2 \pi i}  S \left(   \ln  \frac{\Lambda_0^3}{S} + 1 \right) + \frac{i}{g_s} K S + \sqrt{c''} \frac{S^{2/3}}{\alpha' \sqrt{g_s} M } X \,. \label{4d3}
\end{equation}
The field $X$ satisfies the nilpotent constraint $X^2 = 0$. It contains the goldstino $G$ localized on the antibrane and generates the Volkov-Akulov
nonlinear supersymmetric Lagrangian.  The solution of the constraint, in superspace language is
\begin{equation}
X = \frac{GG}{2 F_X} + \sqrt{2} \theta G + F_X \theta^2 \ , \label{4d4}
\end{equation}
where $\theta$ is the fermionic superspace coordinate.  
The simplest string theory proposal for the realization of the Volkov-Akulov action consists of putting a stuck $\overline{D3}$ antibrane on top of an $O3_{-}$ plane \cite{uranga}, which reduces the (anti)brane localized degrees of freedom to only the goldstino \cite{kallosh}. Similar construction, much like the original string vacua with ``brane 
supersymmetry breaking"  \cite{bsb} also generate a nonlinear realization of supersymmetry on the antibranes, as shown explicitly in \cite{dm}. 
The nilpotent constraint eliminates the scalar partner of the goldstino, keeping the auxiliary field $F_X$. Consequently, the scalar potential  is computed from the usual supergravity potential, by setting at the end $X = 0$. The last term in the superpotential reproduces the   antibrane uplift, redshifted by the
S-dependent prefactor.  

Note that the nilpotent goldstino formalism is valid as long as $F_X \not=0$. In the example we consider
(\ref{4d3}),  we find that $F_X =  \frac{\sqrt{c''}  S^{2/3}}{\alpha' \sqrt{g_s} M }$ and, since $\langle S \rangle \not=0$,  the formalism is indeed valid. The stronger the warping the smaller the supersymmetry
breaking. We expect in principle a maximum value of the warping also  from the requirement that states decoupled by the supersymmetry breaking to be heavy enough.  

We work in the small complex structure limit $S \ll 1$. Taking into account the fact that one sets $X=0$ at the end, the supergravity scalar potential can be approximated by
\begin{equation}
V_{SUGRA} = \kappa^2 _4 e^K \left[ G^{i \bar \j} D_i W \overline{}D_{\bar \j} \bar W - 3 |W|^2 \right] \simeq \kappa^2 _4  
e^{K_0} \left[ G^{S \bar S} |\partial_S W|^2   +  G^{X \bar X} |\partial_X W|^2 \right]  \ . \label{4d5}
\end{equation}
The resulting expression matches correctly the scalar potential found in the previous sections
\begin{equation}
V_{SUGRA} = V_{KS} + V_{\overline D3} \ . \label{4d6}
\end{equation}
Adding the nonperturbative term generated by stringy instantons or gaugino condensation amounts to adding to the perturbative superpotential in (\ref{4d3})
the nonperturbative term $W_{np} = A e^{i a \rho}$, and it is easy to see that this does not affect the conifold destabilization mechanism we found.

\providecommand{\href}[2]{#2}\begingroup\raggedright\endgroup

\end{document}